\documentclass[aps,pra,showpacs,superscriptaddress]{revtex4}
\usepackage{graphicx}
\usepackage{bm}
\usepackage{amsmath}
\newcommand{\fo}{^{\scriptscriptstyle (+)}}
\newcommand{\ba}{^{\scriptscriptstyle (-)}}
\newcommand{\foba}{^{\scriptscriptstyle (\pm )}}
\begin{document}
\date{\today}
\title{Quasiparticle properties in a density functional framework}
\author{D. Van Neck, S. Verdonck, G. Bonny}
\affiliation{Laboratory of Theoretical Physics, Ghent University, 
Proeftuinstraat 86, B-9000 Gent, Belgium}
\author{P.W. Ayers} 
\affiliation{Department of Chemistry, McMaster University, Hamilton, 
Ontario, Canada L8S4M1}
\author{M. Waroquier}
\affiliation{Laboratory of Theoretical Physics, Ghent University, 
Proeftuinstraat 86, B-9000 Gent, Belgium}
\begin{abstract}
We propose a framework to construct the ground-state energy and density 
matrix of an $N$-electron system by solving selfconsistently a set of 
single-particle equations. The method can be viewed as a non-trivial 
extension of the Kohn-Sham scheme (which is embedded as a special case). It 
is based on separating the Green's function into a quasi-particle part and 
a background part, and expressing only the background part as a functional of 
the density matrix. The calculated single-particle energies and wave 
functions have a clear physical interpretation as quasiparticle energies and 
orbitals. 
\end{abstract}
\maketitle

\section{Introduction\label{section1}}

The power of the Kohn-Sham (KS) implementation~\cite{dft} 
of density functional theory (DFT~\cite{Hohen64}) lies in its ability to 
incorporate complex many-body correlations (beyond Hartree-Fock) 
in a computational framework that is not any more difficult 
than the Hartree-Fock (HF) equations. In practice this means one has to 
solve single-particle Schr\"{o}dinger equations, with local or non-local 
potentials, in an iterative self-consistency loop. 
This simplicity makes KS-DFT the only feasible approach in many modern 
applications of electronic structure theory.  
There is therefore continuing interest, not only in developing new
and more accurate functionals, but also in studying conceptual improvements
and extensions to the DFT framework. 

The present implementations of 
DFT can handle short-range interelectronic correlations quite well,
but often fail when dealing with near-degenerate systems characterized
by a small particle-hole gap. This seems to indicate that  
KS-DFT does not accurately describe the Fermi surface if it deviates 
significantly  from the noninteracting 
one~\cite{Zha98,Gri00,Bec03,Gra05,Mol02,Coh01,Han01}. 
In this respect, one of the most glaring inadequacies of KS-DFT is the fact 
that the 
physically important concept of quasiparticles is missing. 
Even formal knowledge of the exact exchange-correlation energy functional 
would only lead to the total energy and the density, since the individual 
KS orbitals have no special significance. An exception is the HOMO orbital 
and energy which govern the asymptotic tail of the density. 
It has also been observed numerically that in exact KS-DFT the occupied 
single-particle energies resemble the exact ionization energies also 
for the deeper valence hole states~\cite{Sav98,Cho02}; the deviations 
have been analyzed theoretically in terms of a reponse contribution 
to the KS potential~\cite{Cho02,Gri03}.   

Quasiparticle (QP) excitations in the Landau-Migdal sense\cite{Lan,Mig} 
form a well-known concept in many-body physics. They are most readily 
understood as the relics of the single-particle (s.p.) excitations in the 
noninteracting system when the interaction is turned on~\cite{Fet71,Dick05}. 
In most electronic systems, or more generally in all normal Fermi systems, 
the bulk of the s.p.\ strength (i.e. the transition strength related 
to the removal or addition of a particle) is concentrated
in QP states. Especially near the Fermi surface, the 
QP states represent the dominant physical feature   
and should be described properly in any appropriate single-particle theory.

The QP orbitals form a complete, linear independent, but generally 
nonorthogonal set. 
The completeness and linear independence follows from the fact that the 
QP states evolve from the set of single-particle eigenstates of a 
noninteracting Hamiltonian. Near the Fermi surface they coincide with the 
electron attachment states or with the dominant ionization states. 
Further away from the Fermi surface, QP states may acquire a width and 
correspond not to a single state but to a group of states in the 
$(N\pm 1)$-electron system characterized by rather 
pure one-hole or one-particle structure.    
The QP states have reduced s.p.\ strength (i.e.\ the normalization of 
the QP orbitals is less than unity). 
Note that QP orbitals and strengths (at least near the Fermi surface) 
are experimentally accessible using, e.g., electron momentum 
spectroscopy~\cite{ems}.  
 
It is the aim of the present paper to develop formally exact 
single-particle equations whose solutions can be interpreted as the  
QP energies and orbitals, and to explore how approximations can be 
introduced by the modelling of small quantities in terms of 
functionals. The resulting formalism will be called QP-DFT and yields, 
apart from the QP orbitals and energies, also the total energy and the density 
matrix of the system. 

To the best of our knowledge this is the first time such a formalism, aiming 
directly at the quasiparticle properties, is proposed. There are some 
similarities to the ``correlated one-particle theory'' as developed recently 
by Beste et al.~\cite{Bes04,Bes05}, 
in that an energy-independent single-particle Hamiltonian 
is obtained whose eigenvalues are related to ionization energies and 
electron affinities. The interesting proposal in Ref.~\cite{Bes04,Bes05}   
is to construct such a s.p.\ Hamiltonian by means of a Fock-space coupled 
cluster expansion, separately in the $N-1$ and $N+1$ electron system, and  
supplemented by a maximal overlap condition between the reference 
Slater determinant and the exact $N$ electron ground state 
(i.e.\ Brueckner coupled cluster). However, it is not obvious how 
the overcomplete set of Dyson orbitals for the $N-1$ 
electron system  is reduced to $N$ single-particle  
orbitals because the technique seems to 
require that in general precisely $N$ exact eigenstates of the $N-1$ electron 
system are projected out as principal ionization states. This 
may be problematic in cases where the s.p.\ strength is fragmented, or 
when considering the electron gas limit. In our opinion, a general 
formulation should be expressed using the QP states and their 
associated width. 

This paper is organized as follows. 
Sec.~\ref{II} reviews all necessary theoretical ingredients of 
the QP-DFT formalism. 
The natural framework for discussing QP properties is 
the propagator or Green's function (GF) formulation of many-body 
theory~\cite{Fet71,Dick05}. 
A summary of some  
GF results is therefore provided in Secs.~\ref{IIA}-\ref{IIB}. 
See, e.g., \cite{Lin73,Ced77,Onid02,Ort95,Hol90} for reviews and
general papers on the specific use of GF theory in electronic structure 
problems. In Sec.~\ref{IIC} the standard quasiparticle concept is introduced. 
The fundamental 
equations of the new QP-DFT formalism are established in Sec.~\ref{IIIA}. 
A discussion of the 
main properties of QP-DFT is then provided in the remainder of 
Sec.~\ref{III}, i.e.\ the embedding of both HF and KS-DFT, the potential 
exactness of the method, the 
asymptotic properties in coordinate space, and the electron gas limit.  
The final Sec.~\ref{IV} contains further discussion of QP-DFT (possible 
implementations, advantages and drawbacks), and concludes with a 
summary of the paper.

\section{Green's function theory and the quasiparticle concept\label{II}}

In this section some well-established results and concepts are reviewed, 
in order to introduce notation and to provide an overview of the 
expressions needed for the QP-DFT formalism.  
The basic relations in GF theory can be found in many textbooks, 
e.g.~\cite{Fet71,Dick05}. In Sec.~\ref{IIC} the quasiparticle 
concept as introduced in~\cite{Lan,Mig} is discussed; this material is also 
extensively treated in~\cite{Dick05,Noz97,Mahaux}.     

\subsection{Single-particle propagator\label{IIA}}

We initially keep the discussion as general as possible, and consider 
a normal (non-superconducting) Fermi system with Hamiltonian 
$\hat{H}=\hat{H}_{0}+\hat{V}$, where $\hat{H}_{0}$ contains the 
kinetic energy and external potential and $\hat{V}$ is the two-particle 
interaction. The s.p.\ propagator in the energy representation 
is defined as 
\begin{equation}
G(\alpha ,\beta ;E)
=\langle\Psi_{0}^{N}|a_{\alpha}\frac{1}{E-(\hat{H}-E_{0}^{N})
+i\eta}a_{\beta}^{\dagger}+a_{\beta}^{\dagger}
\frac{1}{E+(\hat{H}-E_{0}^{N})-i\eta}a_{\alpha}|\Psi_{0}^{N}\rangle\label{eq1}
\end{equation}
where $\alpha,\beta,..$ label the elements of a complete orthonormal
basis set of s.p.\ states, the second-quantization operators $a_{\alpha}$
($a_{\beta}^{\dagger}$) remove (add) a particle in state $\alpha$
($\beta$), and $\eta>0$ is an infinitesimal convergence parameter.
The exact ground state of the $N$-particle system is denoted 
by $|\Psi_{0}^{N}\rangle$ and its energy by $E_{0}^{N}$. 

Inserting complete sets of eigenstates of $\hat{H}$ into Eq.~(\ref{eq1}) 
leads to the standard Lehmann representation of the s.p.\ propagator, 
\begin{eqnarray}
G(\alpha,\beta;E)&=&
\sum_{n}\frac{\langle\Psi_{0}^{N}|a_{\alpha}|\Psi_{n}^{N+1}\rangle
\langle\Psi_{n}^{N+1}|a_{\beta}^{\dagger}|\Psi_{0}^{N}\rangle}
{E-(E_{n}^{N+1}-E_{0}^{N})+i\eta}
+\sum_{n}\frac{\langle\Psi_{0}^{N}|a_{\beta}^{\dagger}|\Psi_{n}^{N-1}\rangle
\langle\Psi_{n}^{N-1}|a_{\alpha}|\Psi_{0}^{N}\rangle}
{E+(E_{n}^{N-1}-E_{0}^{N})-i\eta}\nonumber\\
&\equiv&\sum_{n}\frac{(z_n\fo)_\alpha (z_n\fo)^*_\beta}
{E-\epsilon\fo_n +i\eta}
+\sum_{n}\frac{(z_n\ba)_\alpha (z_n\ba)^*_\beta}
{E-\epsilon\ba_n -i\eta},
\label{eq3}
\end{eqnarray} 
where the $|\Psi_{n}^{N\pm 1}\rangle$ are
the eigenstates, and $E_{n}^{N\pm 1}$ the eigenenergies,
in the $(N\pm1)$-particle system. 
The notation in the last line of Eq.~(\ref{eq3}), i.e.\ 
$\epsilon\foba_n = \pm (E_n^{N\pm 1}-E_0^N)$ for the poles of the 
propagator, and 
\begin{equation}
(z\fo_n )_{\alpha} = \langle\Psi_{0}^{N}|a_{\alpha}|\Psi_{n}^{N+1}\rangle 
;\;\;\; 
(z\ba_n )_{\alpha} = \langle\Psi_{n}^{N-1}|a_{\alpha}|\Psi_{0}^{N}\rangle 
\end{equation}
for the s.p. transition amplitudes, will be used throughout the paper. 
Note that the amplitudes 
$z\foba_n$ are usually 
called Dyson orbitals in the electronic context.  

The poles $\epsilon\fo_n$ and $\epsilon\ba_n$ of the propagator are 
located in the addition domain $(\epsilon_0\fo,+\infty )$ and the 
removal domain $(-\infty , \epsilon_0\ba)$, respectively. 
In a finite system both domains are separated by an energy interval 
$(\epsilon_0\ba ,\epsilon_0\fo )$. The width of the interval is the 
particle-hole gap, 
\begin{equation}
\epsilon_0\fo -\epsilon_0\ba = 
E^{N+1}_0 -2E^N_0 + E^{N-1}_0 >0,\label{eq4}
\end{equation}
where positivity is guaranteed by the assumed convexity of the $E^N_0$ 
versus $N$ curve. Only the interval is physically relevant, 
but for definiteness one can 
take the Fermi energy as the center of the interval,  
\begin{equation}
\epsilon_F = \frac{1}{2}(\epsilon_0\ba+\epsilon_0\fo)
=\frac{1}{2}(E^{N+1}_0 - E^{N-1}_0). 
\end{equation}
In an infinite system one has 
$\epsilon_0\ba=\epsilon_0\fo=\epsilon_F $.

The (1-body) density matrix $[N\ba]$ and removal energy matrix $[M\ba]$ 
can be expressed in terms of the propagator as 
\begin{eqnarray}
N_{\alpha,\beta}\ba & \equiv & 
\langle\Psi_{0}^{N}|a_{\beta}^{\dagger}
a_{\alpha}|\Psi_{0}^{N}\rangle
=\sum_n (z_n\ba)_\alpha (z_n\ba)^*_\beta 
=
\int\frac{dE}{2\pi i}
\mbox{e}^{i\eta E}G(\alpha,\beta;E),\nonumber \\
M_{\alpha,\beta}\ba & \equiv & 
\langle\Psi_{0}^{N}|a_{\beta}^{\dagger}
[a_{\alpha},\hat{H}]|\Psi_{0}^{N}\rangle
= \sum_n \epsilon\ba_n (z_n\ba)_\alpha (z_n\ba)^*_\beta 
=\int\frac{dE}{2\pi i}
\mbox{e}^{i\eta E}EG(\alpha,\beta;E).
\label{eq6}
\end{eqnarray}
The expressions in terms of the $\epsilon_n\ba$ and 
$z\ba_n$ follow by inserting 
the complete set of $(N-1)$-particle eigenstates of $\hat{H}$ between the  
$a_{\beta}^{\dagger}$ and $a_{\alpha}$ operators. The equalities on the 
right of Eq.~(\ref{eq6}) can then be obtained by contour integration in 
the complex $E$-plane, where the $\mbox{e}^{i\eta E}$ factor selects 
the pole contributions in the removal domain.    

Any one-body observable of 
interest can be calculated with the density matrix.
The removal energy matrix allows in addition to calculate the 
total energy through the Migdal-Galitskii sum rule
\begin{equation}
E_{0}^{N}=\frac{1}{2}\mbox{Trace}([H_{0}][N\ba]+[M\ba]),
\label{eq7}
\end{equation}
which can be obtained by exploiting the fact that 
$\mbox{Trace}[M\ba]=\langle\Psi_0^N |\hat{H}_{0}+2\hat{V}|\Psi_0^N\rangle$. 
The removal part of the propagator is sufficient for these purposes. 
However, only the (inverse of) the full propagator has a meaningful 
perturbative expansion, which takes the form of the Dyson equation,  
\begin{equation}
[G(E)]^{-1}=[G_0 (E)]^{-1} - [\Sigma(E)], 
\label{eq8}
\end{equation}
where $[G_0 (E)]$ is the noninteracting propagator corresponding to the 
Hamiltonian $\hat{H}_0$ and $[\Sigma (E)]$ is the (one fermion line) 
irreducible selfenergy. In an ab-initio calculation, the physics input is 
controlled  
by taking a suitable approximation for the selfenergy, but here the reasoning 
is in terms of the exact selfenergy. The latter plays the role 
of an energy-dependent s.p. potential. For a discrete pole of the 
propagator, e.g., the amplitude obeys 
\begin{equation}
([H_0] + [\Sigma (\epsilon\foba_n)])z\foba_n =\epsilon\foba_n z\foba_n ,  
\label{eq9}
\end{equation} 

In Fig.~\ref{fig1} the general diagrammatic structure of the selfenergy is 
shown. The two distinct types also correspond respectively to the sum of all 
energy-independent, and all energy-dependent, contributions to the 
selfenergy. 

\subsection{Spectral function\label{IIB}}

In general, the single-particle spectral function  is related to the 
propagator as 
\begin{eqnarray}
[S(E)]&=&\frac{1}{2\pi i}\mbox{sign}(\epsilon_{F}-E)
([G(E)]-[G(E)]^{\dagger})\label{eq10a}\\
&=&\sum_{n}(z_n\fo)(z_n\fo)^\dagger\delta (E-\epsilon\fo_n)
+\sum_{n}(z_n\ba)(z_n\ba)^\dagger\delta (E-\epsilon\ba_n).
\label{eq10b}
\end{eqnarray}

The zero'th and first energy-weighted moments obey the sum rules 
\begin{eqnarray}
N_{\alpha,\beta} & = & \int_{-\infty}^{+\infty}dES(\alpha,\beta;E)
=\langle\Psi_{0}^{N}|\{ a_{\beta}^{\dagger},a_{\alpha}\}|\Psi_{0}^{N}\rangle
\nonumber \\
M_{\alpha,\beta} & = & \int_{-\infty}^{+\infty}dEES(\alpha,\beta;E)
=\langle\Psi_{0}^{N}|\{ a_{\beta}^{\dagger},
[a_{\alpha},\hat{H}]\}|\Psi_{0}^{N}\rangle,
\label{eq11}
\end{eqnarray}
where the braces denote an anticommutator, e.g. 
$\{a^{\dagger}_{\beta},a_{\alpha}\}=\delta_{\alpha ,\beta}$.  
The equalities on the right of Eq.~(\ref{eq11}) follow [similar  
to Eq.~(\ref{eq6})] by inserting complete sets of eigenstates 
of the Hamiltonian $\hat{H}$ in both the $(N+1)$ and $(N-1)$ system,  
\begin{eqnarray}
\langle\Psi_{0}^{N}|\{ a_{\beta}^{\dagger},
[a_{\alpha},\hat{H}]\}|\Psi_{0}^{N}\rangle &=& 
\langle\Psi_{0}^{N}|a_{\beta}^{\dagger}(E_0^N-\hat{H})
a_{\alpha}+a_{\alpha}(\hat{H}-E_0^N)a_{\beta}^{\dagger}
|\Psi_{0}^{N}\rangle \nonumber\\ 
&=&\sum_{n}\epsilon\fo_n (z_n\fo)(z_n\fo)^\dagger
+\sum_{n}\epsilon\ba_n (z_n\ba)(z_n\ba)^\dagger .
\end{eqnarray} 

Note that the integrations in Eq.~(\ref{eq11}) are over the entire
energy axis. If the integration is restricted to the removal domain
$(-\infty,\epsilon_{F})$ one retrieves the one-body density matrix
$[N\ba]$ and removal energy matrix $[M\ba]$ as defined in
Eq.~(\ref{eq6}), i.e.\
\begin{equation}
N_{\alpha,\beta}=\int_{-\infty}^{\epsilon_{F}}dES(\alpha,\beta;E)
+\int_{\epsilon_{F}}^{+\infty}dES(\alpha,\beta;E)
=N_{\alpha,\beta}\ba+N_{\alpha,\beta}\fo,
\label{eq12}
\end{equation}
and similarly for $M_{\alpha,\beta}=M\ba_{\alpha,\beta}+M\fo_{\alpha,\beta}$.

\begin{figure}
\includegraphics[scale=0.6]{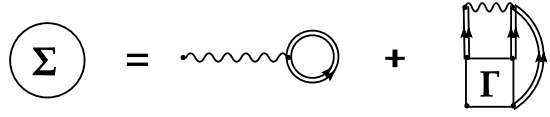}
\caption[fig1]{General structure of the exact self-energy $\Sigma (E)$ in 
terms of the exact 4-point vertex function $\Gamma$. The wavy line represents 
the antisymmetrized interaction, the double directed line is the exact 
propagator. The first term is the sum of all static selfenergy contributions 
$\tilde{V}_{HF}$, the second term involving $\Gamma$ groups all 
energy-dependent contributions.   
\label{fig1}}
\end{figure}

Writing the Hamiltonian as 
\begin{equation} 
\hat{H} = \sum_{\gamma\delta}\langle\gamma|H_0|\delta\rangle 
a^{\dagger}_{\gamma}a_{\delta} 
+\frac{1}{4}
\sum_{\mu\nu\gamma\delta}\langle\mu\gamma|V|\nu\delta\rangle_{as}\, 
a^{\dagger}_{\mu}a^{\dagger}_{\gamma}a_{\delta}a_{\nu},
\end{equation}
with 
\begin{equation}
\langle\alpha\beta|V|\gamma\delta\rangle_{as}
\equiv\langle\alpha\beta|V|\gamma\delta\rangle -
\langle\alpha\beta|V|\delta\gamma\rangle
\end{equation}
the antisymmetrized 
interaction matrix element, 
the (anti)commutator on the right of Eq.~(\ref{eq11}) can be worked out 
explicitly as 
\begin{equation}
\{ a_{\beta}^{\dagger},
[a_{\alpha},\hat{H}]\}
=
\langle\alpha|H_{0}|\beta\rangle
+\sum_{\gamma\delta}\langle\alpha\gamma|V|\beta\delta\rangle_{as}
a^{\dagger}_{\gamma}a_{\delta}.
\end{equation}
As a result, it is possible to express the sumrules in Eq.~(\ref{eq11}) 
in closed form as 
\begin{eqnarray}
N_{\alpha,\beta} & = & \delta_{\alpha,\beta}\mbox{, or }\;\;[N]=[I], 
\label{eq13}\\
M_{\alpha,\beta} & = & \langle\alpha|H_{0}|\beta\rangle
+\sum_{\gamma\delta}\langle\alpha\gamma|V|\beta\delta\rangle_{as}
N_{\delta\gamma}\ba\mbox{, or }\;\;[M]=[H_0]+[\tilde{V}_{HF}],\label{eq14}
\end{eqnarray}
where $[I]$ is the identity matrix. These expressions will be used extensively 
in the following. 
The second term, $[\tilde{V}_{HF}]$, in Eq.~(\ref{eq14}) is the sum of all 
static (energy independent) 
selfenergy contributions, and has the form of the HF mean field, but 
evaluated with the exact density matrix $[N\ba ]$. 
A diagrammatical representation is provided by the first term 
in Fig.~\ref{fig1}. 

\subsection{Quasiparticles\label{IIC}}

In normal Fermi systems, the bulk of the spectral strength is concentrated
in quasiparticle (QP) states which, in the Landau-Migdal picture, 
evolve adiabatically from the $N\pm 1$ eigenstates of $\hat{H}_0$, and 
can be regarded as the elementary s.p.\ excitations in the interacting system. 
In its simplest form the QP contribution to the propagator can be written as 
a modified noninteracting propagator, 
\begin{equation}
G_Q (\alpha, \beta ;E)=\sum_{j=1}^N \frac{(z_{Qj})_\alpha 
(z_{Qj})^*_\beta}
{E-\epsilon_{Qj} -iw_{Qj}} 
+\sum_{j=N+1}^\infty \frac{(z_{Qj})_\alpha (z_{Qj})^*_\beta}
{E-\epsilon_{Qj} +iw_{Qj}}
\label{eq15}
\end{equation} 
where $w_{Qj} >0$ characterizes the width of the QP excitation at energy 
$\epsilon_{Qj}$, and 
$z_{Qj}$ is the corresponding QP orbital. The first term 
in Eq.~(\ref{eq15}) corresponds to  
excitations in the $(N-1)$-particle system, as indicated 
by the location of the poles in the upper half-plane 
($\mbox{Im}E >0$). It contains the $N$ lowest QP energies 
$\epsilon_{Qj}$ ($j=1,..,N$), for which $\epsilon_{Qj}<\epsilon_F$. 
The higher QP states ($j=N+1,..,\infty$) in the second term of  
Eq.~(\ref{eq15}) correspond to the $(N+1)$ system, as indicated by the 
location of the poles in the lower half-plane ($\mbox{Im}E < 0$), and have 
$\epsilon_{Qj} > \epsilon_F$.

According to Eq.~(\ref{eq10a}), the spectral function  corresponding to the 
QP propagator in Eq.~(\ref{eq15}) reads 
\begin{equation}
[S_Q (E)]= \sum_{j=1}^{\infty} (z_{Qj})(z_{Qj})^\dagger 
{\cal L}_{w_{Qj}}(E-\epsilon_{Qj}), 
\label{eq15bis}  
\end{equation}
where the (Breit-Wigner) distribution 
${\cal L}_\Delta (x) = \frac{\Delta /\pi}{x^2 +\Delta^2}$ is 
normalized to unity and has width $\Delta$. 

The set of QP orbitals is complete and linearly independent (in view of 
the 1-1 correspondence to the noninteracting system), but in general 
not orthogonal.  
The QP orbitals  are different from the HF ones because of the 
inclusion of selfenergy contributions beyond HF 
(see second term in Fig.~\ref{fig1}). 
This energy dependent part of the selfenergy is also 
responsible for a reduction of the spectroscopic strength, 
$z_{Qj}^{\dagger}z_{Qj}\leq 1$.      
 
For infinite and homogenuous systems (e.g.\ the electron gas) the 
situation is particularly transparant, since all 
matrix quantities are diagonal in a plane-wave basis, and reduce to continuous 
functions of s.p.\ momentum $p$. 
For momenta close to the Fermi momentum ($p\approx p_F$) the QP energy 
is the solution of 
$\epsilon_Q (p)= p^2/(2m) +\mbox{Re}\Sigma (p; \epsilon_Q (p))$, 
and coincides with the Fermi energy at $p_F$, i.e.\ 
$\epsilon_Q (p_F) =\epsilon_F$.  
The spectral function $S(p;E)$ is dominated by a sharp QP peak 
at $\epsilon_Q (p)$, having a width $w_Q (p)$. 
The width $w_Q (p)$ is proportional 
to $\mbox{Im}\Sigma (p;\epsilon_Q (p))$  and vanishes for $p\rightarrow p_F$. 
While the QP excitations are only unambiguously defined near the Fermi 
surface, the concept can be extended to all momenta. 

Similar considerations also apply to finite systems, where the QP excitation 
corresponds to a state or group of states having rather pure s.p. character. 
The $(N\pm 1)$-particle states near the Fermi energy (where no poles of the 
selfenergy are present) are usually characterized by spectroscopic 
factors of the order of (but smaller than) unity.  
In closed-shell atoms, e.g., the first ionization state
$\Psi_{0}^{N-1}$ (corresponding to the first pole in the removal
domain of the propagator) typically has a spectroscopic factor of
about 95\% \cite{Neck01}. In that case the QP orbital in Eq.~(\ref{eq15}) 
coincides with the Dyson orbital. More deeply bound orbitals may acquire a 
width, but have similar summed strength concentrated near an average 
QP energy. The QP orbital in Eq.~(\ref{eq15}) is then representative for the 
transition amplitudes to these states.  

The QP contribution to the sumrules in Eq.~(\ref{eq11}) is given by 
\begin{equation}
[N_Q] = \sum_{j=1}^\infty z_{Qj}z_{Qj}^{\dagger};\;\;\;
[M_Q] = \sum_{j=1}^\infty \epsilon_{Qj} z_{Qj}z_{Qj}^{\dagger}.
\label{eq16}
\end{equation}
Note that the QP width does not contribute to the $0^{\mbox{th}}$ and 
$1^{\mbox{st}}$ moment, and 
drops out from all subsequent considerations.

Finally, it should be noted that the simple ansatz in Eq.~(\ref{eq15}) 
does not provide a valid description of a QP excitation $j$
far away from the QP energy $\epsilon_{Qj}$.  
The energy-dependence in Eq.~(\ref{eq15bis}) 
can then be slightly misleading, as the tails of the Breit-Wigner 
distribution extend beyond the Fermi energy. 
In reality the strength of a QP excitation either belongs to the $N-1$ or 
to the $N+1$ system. The QP contribution to the separate 
$[N\foba]$ and $[M\foba]$ 
should therefore not be determined 
directly by Eqs.~(\ref{eq12},\ref{eq15bis}), but rather 
by the location of the poles in the lower or 
upper halfplane, i.e. 
\begin{eqnarray}
[N_Q\ba]&=&\int \frac{dE}{2\pi i}\mbox{e}^{i\eta E} [G_Q (E)] 
=\sum_{j=1}^N z_{Qj}z_{Qj}^{\dagger}\nonumber\\
\mbox{}[M_Q\ba]&=&\int \frac{dE}{2\pi i}\mbox{e}^{i\eta E} E[G_Q (E)] 
=\sum_{j=1}^N \epsilon_{Qj} z_{Qj}z_{Qj}^{\dagger}, 
\label{eq17bis}
\end{eqnarray}
and similarly for 
\begin{equation}
[N_Q\fo]=\sum_{j=N+1}^\infty z_{Qj}z_{Qj}^{\dagger};\;\;\;
[M_Q\fo]=\sum_{j=N+1}^\infty \epsilon_{Qj} z_{Qj}z_{Qj}^{\dagger}. 
\label{eq17ter}
\end{equation}

\section{Quasiparticle equations\label{III}}

\subsection{Derivation of the QP-DFT equations\label{IIIA}}

For the following it is important to realize that, given 
arbitrary hermitian matrices $[N_Q]$ and $[M_Q]$ with $[N_Q]$  
positive-definite, 
one can always write the unique decomposition of Eq.~(\ref{eq16}). 
This can be achieved by constructing the unique basis that solves 
the (generalized) eigenvalue problem 
\begin{equation} 
[M_Q]u_j = \lambda_j [N_Q]u_j;\;\;\;\;u_j^\dagger [N_Q] u_k =\delta_{j,k}, 
\label{eq17}
\end{equation}
where $[N_Q]$ plays the role of a metric matrix; the QP energies and 
orbitals given by $\epsilon_{Qj}=\lambda_j$ and $z_{Qj}=[N_Q]u_j$ then 
fulfill Eq.~(\ref{eq16}). 

The eigenvalue problem  in Eq.~(\ref{eq17}) can be considered as a 
set of s.p. equations determining the QP orbitals and energies. We now    
rewrite the unknown operators $[N_Q]$ and $[M_Q]$ in a more useful form 
that suggests possible approximation 
schemes.  

Since the QP contribution to the spectral function is dominant, it makes sense 
to isolate it, 
\begin{equation}
[S(E)]=[S_{Q}(E)]+[S_{B}(E)],
\label{eq19}
\end{equation}
and concentrate on the residual small 'background' contribution $[S_B (E)]$.
In fact, the full energy dependence of $[S_{B}(E)]$ is not needed,
since one can apply the quasiparticle-background separation of 
Eq.~(\ref{eq19}) to the zero'th and first energy-weighted moments as well, 
i.e.\ one has
\begin{equation}
[N]=[N_{Q}]+[N_{B}];\;\;\;[M]=[M_{Q}]+[M_{B}],
\label{eq20}
\end{equation}
where the total energy integrals $[N_B]=[N_B\ba]+[N_B\fo]$ and 
$[M_B]=[M_B\ba]+[M_B\fo]$ can again be split  
in a removal and addition part. 
Note that the matrices $[N]$ and $[M]$ on the left side of 
Eq.~(\ref{eq20}) are known in closed form through 
Eqs.~(\ref{eq13}-\ref{eq14}), so it follows that 
\begin{eqnarray}
[N_Q]&=&[I]-[N_B],\label{neweq1}\\
\mbox{}[ M_Q ]&=&[H_0]+[\tilde{V}_{HF}]-[M_B].\label{neweq2}
\end{eqnarray}
 
One then arrives at the remarkable conclusion that modelling the background 
contributions $[M_{B}\foba]$
and $[N_{B}\foba]$ as a functional of e.g.\ the density matrix $[N\ba]$, 
is sufficient to generate a selfconsistent set of s.p. equations.
Using Eqs.~(\ref{neweq1}-\ref{neweq2}) the eigenvalue problem 
in Eq.~(\ref{eq17}) can be expressed as 
\begin{equation}
([H_0]+[\tilde{V}_{HF}\{N\ba\}]-[M_{B}\{N\ba\}])u_j=
\lambda_j ([I]-[N_{B}\{N\ba\}])u_j ,
\label{eq21}
\end{equation}
where the functional dependency of $[M_B]$ and $[N_B]$ is indicated 
between braces. Note that also the HF-like potential $[\tilde{V}_{HF}]$  
is by definition [see Eq.~(\ref{eq14})] expressed in terms of the density 
matrix, as  indicated in Eq.~(\ref{eq21}).   

Having an initial estimate for $[N\ba]$ allows to 
construct the matrices $[N_B]$ and $[M_B]$, as well as the HF-like 
potential $[\tilde{V}_{HF}]$.
The eigenvalue problem in Eq.~(\ref{eq21}) can now be solved, yielding 
QP energies $\epsilon_{Qj}=\lambda_j$ and orbitals 
$z_{Qj}=([I]-[N_{B}\{N\ba\}])u_j$. 
The $N$ solutions with lowest energy represent excitations in the
$N-1$ system, and should be used to update the density matrix 
$[N\ba]=[N_Q\ba]+[N_B\ba]$, 
\begin{equation}   
[N\ba_{\mbox{\small new}}]=\sum_{j=1}^N z_{Qj}z_{Qj}^{\dagger} 
+[N_B\ba\{N\ba\}].
\label{eq22}
\end{equation}
This closes the selfconsistency loop, which can be iterated 
to convergence. The total energy then follows from Eq.~(\ref{eq7}), 
\begin{equation}
E_0^N = \frac{1}{2}\sum_{j=1}^N z^{\dagger}_{Qj}([H_0]+\epsilon_{Qj})z_{Qj} 
+\frac{1}{2}\mbox{Trace}\left([H_{0}][N\ba_B\{N\ba\} ]
+[M_B\ba\{N\ba\}]\right).
\label{eq23}
\end{equation}

The above formalism, henceforth called quasiparticle DFT (QP-DFT), 
generates the total energy, the density matrix, 
and the individual QP energies and orbitals, starting from a model 
for the background contributions $[M_{B}\foba]$ and $[N_{B}\foba]$ 
as a functional of the density matrix. It is intuitively clear 
that this is a reasonable strategy: the external potential 
appears directly in the QP hamiltonian $[M]-[M_B]$ through the 
$[H_0]$ term of Eq.~(\ref{eq14}), and primarily influences the position of 
the QP peaks. One may then assume the background part 
to be generated by 'universal' electron-electron correlations, and to be a 
good candidate for modelling.  
Equation~(\ref{eq21}), determining the QP energies and orbitals, is the 
central result of the present paper, and its properties will now 
be discussed in detail.

\subsection{HF and KS-DFT as special cases of QP-DFT\label{HFandKS}}
We first show that both HF and KS-DFT theory are included in the 
general QP-DFT treatment. 

For HF this is rather obviously achieved by setting all background quantities 
$[M_B\foba]$ and $[N_B\foba]$ equal to zero in Eq.~(\ref{eq21}).  
Since the metric matrix on the right of Eq.~(\ref{eq21}) is now simply 
the identity matrix, one has $z_{Qj}=u_j$ and the QP orbitals form an 
orthonormal set obeying  $([H_0]+[\tilde{V}_{HF}])z_{Qj}=\epsilon_{Qj} z_{Qj}$.
The matrix $[\tilde{V}_{HF}]$ is given by Eq.~(\ref{eq14}) where 
the density matrix $[N\ba]$ in the present approximation follows from 
Eq.~(\ref{eq22}) with $[N\ba_B]=0$, i.e. 
\begin{equation}   
[N\ba]=\sum_{j=1}^N z_{Qj}z_{Qj}^{\dagger}. 
\label{eq24}
\end{equation} 
One can see that $[M]=[H_0]+[\tilde{V}_{HF}]$ 
is just the ordinary HF hamiltonian, and the 
total energy obtained from Eq.~(\ref{eq23}) by setting $[N_B\ba]=[M_B\ba]=0$, 
\begin{equation}
E_0^N = \frac{1}{2}\sum_{j=1}^N (z_{Qj}^{\dagger}[H_0]z_{Qj} +
\epsilon_{Qj})  
\end{equation}
is equivalent to the HF total energy. 

The KS-DFT case is somewhat more difficult, but one also starts 
by setting $[N_B\foba]=0$ in Eq.~(\ref{eq21}), so the QP states $z_{Qj}$ 
form an orthonormal set obeying 
\begin{equation}
([H_0]+[\tilde{V}_H]+[\tilde{V}_F]-[M_B])z_{Qj}=
\epsilon_{Qj}z_{Qj}.
\label{QKS1}
\end{equation}
Here the HF-like potential has been split into its direct and exchange 
components, $[\tilde{V}_{HF}]=[\tilde{V}_{H}]+[\tilde{V}_{F}]$. As the 
density matrix is again given by Eq.~(\ref{eq24}), both components  
only receive contributions from the occupied ($j=1,..,N$) orbitals $z_{Qj}$:
\begin{eqnarray}
(\tilde{V}_H )_{\alpha ,\beta}
&=&\mbox{ }\sum_{\gamma\delta}
\langle\alpha\gamma|V|\beta\delta\rangle
\sum_{j=1}^N (z_{Qj})_{\delta}
(z_{Qj})^*_{\gamma},
\nonumber\\
(\tilde{V}_F )_{\alpha ,\beta}&=&
\mbox{}-\sum_{\gamma\delta} 
\langle\alpha\gamma|V|\delta\beta\rangle  
\sum_{j=1}^N (z_{Qj})_{\delta}
(z_{Qj})^*_{\gamma} .
\end{eqnarray}

For the Coulomb interaction and taking as s.p. labels for the 
electrons the space coordinate and (third component of) 
spin, $\alpha \equiv {\bf r} m_s$ , this reduces to the familiar 
expression
\begin{equation}
\tilde{V}_H ({\bf r}m_{s}, {\bf r}'m'_{s})+
\tilde{V}_F ({\bf r}m_{s}, {\bf r}'m'_{s})=
\delta_{m_s ,m'_s}\delta ({\bf r} -{\bf r}' )
\sum_{j=1}^N \sum_{m''_s}\int d{\bf r}'' \frac{|z_{Qj}({\bf r}''m''_s)|^2}
{|{\bf r}-{\bf r}''|} 
-\sum_{j=1}^N\frac{z_{Qj}({\bf r}m_s)z^*_{Qj}({\bf r}'m'_s)}
{|{\bf r}-{\bf r}'|}. 
\end{equation}
Note that for compactness  we  continue to employ the general matrix 
notation used so far, with the understanding that sums over s.p.\ labels 
should be replaced by coordinate space integrations where appropriate. 
The total energy follows from Eq.~(\ref{eq23}) with $[N\foba_B]=0$,
\begin{equation}
E_0^N = \frac{1}{2}\sum_{j=1}^N (z_{Qj}^{\dagger}[H_0]z_{Qj} +
\epsilon_{Qj})+\frac{1}{2}\mbox{Trace}[M_B\ba].
\label{QKS2}
\end{equation}
The unknown $[M\foba_B]$ should now be determined by 
identification with the results of a KS-DFT calculation. 

We allow explicit dependence on the KS orbitals $\varphi_{KSj}$, 
and write the 
exchange-correlation energy functional as
\begin{equation}
E_{xc}=\sum_{j=1}^N \varphi_{KSj}^{\dagger}[\epsilon_{xc}]\varphi_{KSj}. 
\end{equation}
Specializing e.g.\ to a hybrid 
functional (the derivation for a functional with general 
orbital dependencies proceeds in a similar fashion)  
with a fraction $\beta$ of exact exchange, the matrix $[\epsilon_{xc}]$ in 
coordinate-spin space reads,
\begin{equation}
\epsilon_{xc}({\bf r}\,m_s ,{\bf r}'\,m'_s ) =
-\frac{\beta}{2}\sum_{j=1}^N \frac{
\varphi_{KSj} ({\bf r}\,m_s )\varphi_{KSj}^* ({\bf r}'\,m'_s )}
{|{\bf r}-{\bf r}'|}
+\delta_{m_s ,m'_s }\delta({\bf r}-{\bf r}')f_{\beta}
(\rho ({\bf r}), \nabla\rho ({\bf r})),
\end{equation}
where the second term contains a local functional of the electron density 
$\rho ({\bf r})$ and 
its gradient (assuming for simplicity a spin-saturated system). 
The corresponding exchange-correlation potential $[V_{xc}]$, appearing in the 
KS equation, then reads 
\begin{equation}
V_{xc}({\bf r}\,m_s ,{\bf r}'\,m'_s) =
-\beta\sum_{j=1}^N \frac{
\varphi_{KSj} ({\bf r}\,m_s )\varphi_{KSj}^* ({\bf r}'\,m'_s)}
{|{\bf r}-{\bf r}'|}
+\delta_{m_s ,m'_s }\delta({\bf r}-{\bf r}')\{  f_{\beta}
+\rho \frac{\partial f_\beta }{\partial \rho}
-\nabla\cdot (\rho\frac{\partial f_\beta}{\partial \nabla\rho})\}.
\end{equation}
Identification of the orbitals and energies in the KS equation 
\begin{equation}
([H_0]+[V_H]+[V_{xc}])\varphi_{KSj}=
\epsilon_{KSj}\varphi_{KSj}
\label{KS1}
\end{equation}
with those in Eq.~(\ref{QKS1}) then requires 
\begin{equation}
[M_B]=[V_F]-[V_{xc}]. 
\label{KS1bis}
\end{equation}
Identification of the KS total energy 
\begin{eqnarray}
E_0^N &=& \sum_{j=1}^N \varphi_{KSj}^{\dagger}([H_0]+[V_H])
\varphi_{KSj}+E_{xc}\nonumber\\
&=&
\frac{1}{2}\sum_{j=1}^N (\varphi_{KSj}^{\dagger}[H_0]\varphi_{KSj} +
\epsilon_{KSj})+E_{xc}
-\frac{1}{2}\sum_{j=1}^N \varphi_{KSj}^{\dagger}[V_{xc}]\varphi_{KSj}
\label{KS2}
\end{eqnarray}
with Eq.~(\ref{QKS2}) requires 
\begin{equation}
\mbox{Trace}[M_B\ba]=\sum_{j=1}^N \varphi_{KSj}^{\dagger}
(2[\epsilon_{xc}]-[V_{xc}])\varphi_{KSj}.
\label{KS2bis}
\end{equation}
Choosing the $[M_B\foba]$ operators as  
\begin{equation}
[M_B\ba]= [P](2[\epsilon_{xc}]-[V_{xc}])[P];\;\;\;
[M_B\fo]=[V_F]-[V_{xc}] -  [P](2[\epsilon_{xc}]-[V_{xc}])[P],
\label{KS3}
\end{equation}
where $[P]=\sum_{j=1}^N z_{Qj}z_{Qj}^{\dagger}$ projects onto the 
occupied QP orbitals, fulfills the requirements of both Eq.~(\ref{KS1bis}) 
and of Eq.~(\ref{KS2bis}); this choice therefore leads to the same results as 
the KS-DFT approximation, for the total energy as well as the orbitals 
and orbital energies.  

One concludes that the QP-DFT formulation is flexible enough to reproduce  
HF or KS-DFT results by specific  choices of $[N_B\foba]$ and $[M_B\foba]$. 
Note that the KS-DFT formalism is embedded in a slightly tortuous way, as 
$[N_B\foba]=0$ would normally imply the absence of a background contribution 
to the first-order moment $[M_B\foba]$ as well. 
A similar situation is encountered in 
Brueckner-Hartree-Fock theory\cite{BHF}, 
in which the real part of the selfenergy causes a shift in the QP energy, 
but the corresponding reduction of strength through the imaginary part of the 
selfenergy is neglected. 

\subsection{Potential exactness}

It should be clear that GF quantities like  
the propagator, spectral function, or selfenergy, 
are well-defined and can in principle be calculated 
exactly. Also the separation of the spectral strength into QP and 
background parts can be performed for any system. In fact, 
defining the QP part can usually be done in several ways  
which are all, however, equivalent at the Fermi surface.   
Note that the density matrix and the total energy only contain 
the $0^{\mbox{th}}$ and $1^{\mbox{st}}$ energy-weighted 
moment $[N\ba]$ and $[M\ba]$, and do not depend on the 
QP-background separation. 
As long as one sticks to one unique prescription, the QP 
orbitals and energies are well-defined quantities, and so are the 
background contributions $[N_B\foba]$ and $[M_B\foba]$. 
The fundamental expressions in Eq.~(\ref{eq21}-\ref{eq23}) 
are therefore always valid.

As an example of an unambiguous definition of QP excitations one can 
e.g., in the spirit of Landau-Migdal theory, consider the 
lowest order expansion of the selfenergy around $\epsilon_F$: 
\begin{equation}
[\Sigma (E)]=[\Sigma (\epsilon_F )]+ (E-\epsilon_F)[\Sigma '(\epsilon_F)],
\label{LM}
\end{equation}
where $[\Sigma '(E)]$ is the first derivative of the selfenergy with respect 
to energy. For infinite homogenuous systems the right side of Eq.~(\ref{LM}) 
is a real function, since $\mbox{Im}\Sigma (p, E )
=\frac{\partial}{\partial E}\mbox{Im}\Sigma (p, E )=0$  at the 
Fermi energy $E=\epsilon_F$. 
In a finite system the right of Eq.~(\ref{LM}) is an hermitian matrix, 
since $[\Sigma (E)]=[\Sigma (E)]^{\dagger}$ in a finite pole-free interval 
around the Fermi energy. 
Substituting the linearized selfenergy~(\ref{LM}) into Eq.~(\ref{eq9}) 
leads to 
\begin{equation}
([H_0]+
[\Sigma (\epsilon_F )]-\epsilon_F [\Sigma '(\epsilon_F)])u=
E([I]-[\Sigma '(\epsilon_F)])u ,
\label{LMQP}
\end{equation}
defining the QP excitations. Note the formal similarity to 
Eq.~(\ref{eq21}), and the fact that the metric matrix is positive-definite
as $[\Sigma '(\epsilon_F)]$ is a negative-definite hermitian matrix.  
While Eq.~(\ref{LMQP}) results in a correct description for 
$E\approx\epsilon_F$, it may not be appropriate away from the Fermi surface,   
when the linear approximation~(\ref{LM}) breaks down. 

It should be stressed that the possibility of defining QP excitations in 
various ways is not a shortcoming of the present approach, but rather a 
reflection of the physical reality that QP excitations are only unambiguously 
defined near the Fermi surface, where the density of states in the 
$N\pm 1$ system is small. 
The complete description of the s.p.\ properties in an interacting many-body 
system is contained in the energy dependence of the spectral function. 
{\em Any} effective s.p. Hamiltonian can at most describe the peaks in the 
spectral function, i.e.\ identify the energy regions where the s.p.\  
strength is concentrated and assign an average transition amplitude to this 
region.    

Finally we should explain the use of the density matrix 
as the independent variable for the functional 
modelling of $[N_B\foba]$ and $[M_B\foba]$ in Eq.~(\ref{eq21}). 
In the present context it 
seems most natural to do this in terms of a matrix quantity. 
Moreover, for the class of systems with fixed two-body interaction $\hat{V}$ 
and varying nonlocal external potentials (i.e.\ varying $\hat{H}_0$), 
the fundamental theorem of density matrix functional theory~\cite{dmft}  
implies that all quantities can be expressed as functionals of the density 
matrix. This holds in particular for the $[N_B\foba]$ and $[M_B\foba]$, 
once a unique prescription is agreed upon. 

It is of course also possible to consider the electron density $\rho$ as the 
basic variable, provided one restricts oneself to the class of systems 
with local external potentials. In that case one is again formally assured 
of the existence of functionals for all quantities. Note that the 
functional in Sec.~\ref{HFandKS} reproducing the KS-DFT results, even when 
the exact KS functional is used, is not the 
exact QP-DFT functional since it does not yield the correct density matrix; 
the exact QP-DFT functional would definitely have $[N_B\foba]\neq 0$. 

\subsection{Asymptotics in coordinate space for finite systems}

For notational convenience the spin dependence is dropped here.   
In coordinate space the background quantities $N_B\ba ({\bf r},{\bf r}')$ 
and $M_B\ba ({\bf r},{\bf r}')$  represent nonlocal operators, as they 
have a finite trace. 
The complete background operators $N_B ({\bf r},{\bf r}')$ 
and $M_B ({\bf r},{\bf r}')$ can be local or nonlocal. 
    
It seems plausible that the background operators are short-ranged in the sense 
that correlation effects should become insignificant at large distances.  
I.e.\ the limit $|{\bf r}|\rightarrow\infty $ of the background operators  
should drop out when considering the 
asymptotic behavior of the quasiparticles in Eq.~(\ref{eq21}).  
This requires for $|{\bf r}|\rightarrow\infty $ that 
\begin{equation}
| \int d{\bf r}' N_B({\bf r}, {\bf r}') z({\bf r}') |
/|z({\bf r})|\rightarrow 0 
\label{shra}  
\end{equation}
for an arbitrary s.p.\ orbital $z({\bf r})$. It is sufficient to 
take $[N_B]$ local and asymptotically vanishing, but nonlocal implementations 
are also possible. For $[M_B]$ one requires that for 
$|{\bf r}|\rightarrow\infty $, 
\begin{equation} 
M_B ({\bf r},{\bf r}')\rightarrow -\frac{N_B\ba ({\bf r},{\bf r}')}
{|{\bf r}-{\bf r}'|} + H_B({\bf r},{\bf r}')
\label{shra2}
\end{equation}
where $H_B({\bf r},{\bf r}')$ is again short-ranged in the sense of 
Eq.~(\ref{shra}). Note that one does not require $[N_B\ba]$ to be short-ranged, 
but its contribution to the Fock term in $[\tilde{V}_{HF}\{N\ba\}]$ of 
Eq.~(\ref{eq21}) should be canceled by $[M_B]$ in the limit 
$|{\bf r}|\rightarrow\infty $. The $[N_B\ba ]$ contribution to the Hartree term 
remains, since this is not affected when taking 
the limit $|{\bf r}|\rightarrow\infty $ in Eq.~(\ref{eq21}).  

With the characteristics~(\ref{shra}-\ref{shra2}) of the exact QP-DFT functional, 
the background operators simply drop out of the asymptotic behavior of 
Eq.~(\ref{eq21}), 
and one is left with (for an external charge $Z$)
\begin{equation}
\left(-\frac{\nabla^2}{2m}-\frac{Z-N}{r}-\epsilon_{Qi}\right)z_{Qi}({\bf r})  
-\sum_{j=1}^N z_{Qj}({\bf r})
\int \frac{d{\bf r'}}{|{\bf r}-{\bf r}'|}   
z^*_{Qj}({\bf r}')  u_{i}({\bf r}')
=0.
\label{asymp}
\end{equation}
Note that, for $|{\bf r}|\rightarrow\infty$,  one has 
$z_{Qi}({\bf r})=  u_{i}({\bf r})$ since the metric matrix $[I]-[N_B]$ 
becomes the identity by virtue of Eq.~(\ref{shra}). The Hartree term 
gives rise to the $N/r$ potential, and only the Fock term built with the 
$N$ lowest QP orbitals needs further analysis. In fact, one immediately sees 
that the behavior of the Fock term is the same as the one studied in 
Ref.~\cite{Handy} (or more generally in Ref.~\cite{Kat80}). 
The asymptotic expansion 
\begin{equation}
\frac{1}{{|{\bf r}-{\bf r}'|}}=\frac{1}{r}+\sum_{L=1}^{\infty}
\frac{r'^{L}}{r^{L+1}}P_L (\mbox{cos}\omega ), 
\label{expansion}
\end{equation}
where $P_L$ is a Legendre polynomial and $\omega$ the angle between 
${\bf r}$ and ${\bf r}'$, can be substituted in the Fock term of 
Eq.~(\ref{asymp}).  
The first ($L=0$) term in Eq.~(\ref{expansion}) then  
only contributes in case of removal-type ($i=1,..,N$) 
QP orbitals, since [see Eq.~(\ref{eq17})]
\begin{equation}
\int d{\bf r'}\, z^*_{Qj}({\bf r}')  u_{i}({\bf r}') 
= z^{\dagger}_{Qj} u_i = u^{\dagger}_j [N_Q] u_i = \delta_{i,j},
\end{equation}
and effectively changes the asymptotic charge to $Z-N+1$. 
A similar analysis as in Ref.~\cite{Kat80} then proves that 
the removal QP orbital with the slowest decay is the HOMO one ($i=N$),  
and goes like $\sim \mbox{e}^{-\alpha r}r^\beta$      
with $\alpha^2 =-2\epsilon_{QN}$ and $\beta=(Z-N+1)/\alpha$. 
The Fock term imposes the same exponential decay (except in very special 
cases) on the other removal QP orbitals, but with an accompanying power law 
that decreases faster by at least $1/r^2$. 

For the addition-type ($i=N+1,..,+\infty$) QP orbitals the first term in 
Eq.~(\ref{expansion}) 
does not contribute to the Fock term in Eq.~(\ref{asymp}), and the 
asymptotic charge remains $Z-N$. This means that for neutral atoms 
no Rydberg sequence of 
bound addition-type QP orbitals is present, which is in agreement with the 
real physical situation. Also note that, unlike HF, 
the presence of the $[M_B]$ term in 
Eq.~(\ref{eq21}) does allow for the possibility that {\em some} 
addition-type QP orbitals become bound for neutral systems. 
The asymptotics of the bound addition-type QP orbitals 
is again correct, with an exponential decay governed by their individual 
separation energies $\epsilon_{Qi}$. They go like 
$\sim \mbox{e}^{-\alpha r}r^\beta$      
with $\alpha^2 =-2\epsilon_{Qi}$ and $\beta=(Z-N)/\alpha$; the 
Fock term in Eq.~(\ref{asymp}) cannot influence this, since all 
removal-type QP orbitals decay faster.

\subsection{The electron gas limit}

We consider for simplicity the spin-unpolarized electron gas at 
density $\rho$, and suppress the spin indices. 
To find the electron gas limit of the $[M\foba_B]$ and $[N\foba_B]$ 
quantities one needs the momentum distribution $n(p)$ and the removal 
energy $r(p)$, defined as
\begin{equation} 
n(p)=\langle\Psi_0|a^{\dagger}_{\bf p}a_{\bf p}|\Psi_0\rangle ;\;\;
r(p)=\langle\Psi_0|a^{\dagger}_{\bf p}[a_{\bf p}, \hat{H}]|\Psi_0\rangle . 
\end{equation}
The QP-background separation can be made once a QP spectrum $\epsilon_Q (p)$ 
and strength $Z_Q(p)$ is defined. Near the Fermi surface the standard 
definitions 
\begin{equation}   
\epsilon_Q (p)= p^2/(2m) + \mbox{Re}\Sigma (p; \epsilon_Q (p));\;\;\;
Z_Q (p) = 1/(1-\frac{\partial\mbox{Re}\Sigma (p;E)}
{\partial E})_{E=\epsilon_Q (p)}
\label{standard}
\end{equation}
are valid. For momenta far from $p_F$ Eq.~(\ref{standard}) may break down, 
e.g.\ the equation defining $\epsilon_Q (p)$ may have multiple roots, or the 
strength $Z_Q (p)$ may become larger than unity. An alternative definition, 
valid for all momenta, must then be used. For an example of such a procedure 
we refer to~\cite{Dewu05}, where the QP-background separation was made 
in the context of a $GW$ description~\cite{Hed65,vB&H} of the electron gas.  

The background quantities in the electron gas are now expressed as 
\begin{eqnarray}
N\ba_B (p) &=& n(p) -\theta (p_F -p)Z_Q (p);\;\;\;
N\fo_B (p) = 1-n(p) -\theta (p-p_F )Z_Q (p)\nonumber\\
M\ba_B (p) &=& r(p) -\theta (p_F -p)Z_Q (p)\epsilon_Q (p);\;\;\;
M\fo_B (p) = p^2/(2m)+\tilde{V}_F (p) -r(p) 
-\theta (p-p_F )Z_Q (p)\epsilon_Q (p) ,
\end{eqnarray}
where $\tilde{V}_F (p)$ is the Fock-like potential (evaluated with the 
exact momentum distribution $n(p)$).  
The typical behavior of these background quantities as a function of momentum 
can be found in Fig.~1 of Ref.\cite{Dewu05}. 

For the purpose of applying the electron gas results to inhomogenuous systems, 
it is not a priori clear what the equivalent of the Local Density 
Approximation would be. Working with 
the Wigner transform would seem the most straightforward,  
\begin{eqnarray}
N_B\ba ({\bf r}\,m_s, {\bf r}'\,m'_s )&=& \delta_{m_s ,m'_s}
\left.\int \frac{d{\bf p}}{(2\pi)^3} 
\mbox{e}^{i{\bf p}\cdot({\bf r}-{\bf r}')}
N\ba_{B}(p)\right|_{\rho=\rho({\bf R})}, 
\end{eqnarray}
but there are several possibilities which reduce to 
the same electron gas result in the homogenuous limit.  

\section{Summary and discussion\label{IV}}

The concept of quasiparticles is an important 
tool to understand and describe normal Fermi systems. 
In this paper we developed a set of single-particle equations 
whose solutions correspond to the QP orbitals and energies. 
When the residual small background contributions are expressed 
as universal functionals of the density or density matrix, a single-particle 
selfconsistency problem (the QP-DFT scheme) is generated that can be easily 
solved for an approximate choice of the functionals. 

The QP-DFT scheme would seem to offer many advantages as compared to KS-DFT. 
There is no need for the difference between the kinetic energy of the 
interacting systems and a reference system. 
All s.p. orbitals and energies have  physical meaning, 
in contrast to the KS orbitals. The asymptotic behavior of the QP orbitals 
comes out correct, provided the background operators are short-ranged. 
On the down side: since we no longer have a sharp Fermi surface, 
particle-number conservation is not automatically guaranteed, and should be 
built into the functionals. 
 
The fact that KS-DFT is built in as a special case is a very important 
feature of QP-DFT. In a sense, one cannot do worse than KS-DFT, since 
one adds more parameters to the model. 
Moreover, the new parameters are truly new degrees of freedom 
(the introduction of the metric matrix, allowing a softening of the 
Fermi surface), which cannot 
be mimicked by taking a more sophisticated KS functional.  

The modelling of the background operators $[N_B\foba]$, $[M_B\foba]$ 
is basically virgin territory. 
One option would be to exploit the relation~(\ref{KS3}) 
with the KS formalism, using an existing XC functional form for $[M_B\foba]$, 
adding a similar form for $[N_B\foba]$, and (re)parametrize by fitting to 
total energies and ionization energies in a training  set of atoms and 
molecules.     

Alternatively, 
one could devise parametrizations by performing GF calculations on a 
series of test systems, and construct the background operators directly from 
the calculated spectral function. A step in this direction was taken in  
Ref.~\cite{g0w0}, 
where we applied an ab-initio selfenergy of the $G_0 W_0$ type  
to a series of closed-shell atoms. The QP-DFT scheme outlined in the present 
paper was used in first iteration (no selfconsistency) to generate 
the first-order corrections to the HF picture. We then constructed a 
simple QP-DFT functional that depends only on HF quantities, 
but was able to reproduce the most important results of the underlying 
ab-initio model. 
\begin{acknowledgments}
P.W.A. acknowledges support from NSERC and the Canada Research Chairs, 
S.V. from the Research Council of Ghent University.  
\end{acknowledgments}

\end{document}